\documentstyle[preprint,aps,epsfig]{revtex}
\begin{document}

\preprint{SOGANG-ND 77/98}
\title{Characteristic Relations of Type-III Intermittency in an Electronic
Circuit}
\author{ Chil-Min Kim$^*$ and Geo-Su Yim}
\address{Department of Physics, Pai Chai University, Taejon 302-735, Korea}
\author{Jung-Wan Ryu and Young-Jai Park$^{**}$}
\address{Department of Physics, Sogang University, Seoul 121-742, Korea}
\date{Received 21 January 1988; published 4 May 1988}
\maketitle

\begin{abstract}
It is reported that the characteristic relations of type-II and type-III
intermittencies, with respective local Poincar$\acute{e}$ maps of
$x_{n+1}= (1+\epsilon )x_n + a x^3_n$ and $x_{n+1}= - (1+\epsilon )x_n - a x^3_n$,
are both $- ln (\epsilon)$ under the assumption of uniform reinjection
probability. However, the intermittencies have various characteristic
relations such as $\epsilon^{-\nu}$ ($1/2 \leq \nu \leq 1$) depending on the
reinjection probability. In this Letters the various
characteristic relations are discussed, and the $\epsilon^{-1/2}$
characteristic relation is obtained experimentally in an electronic circuit,
with uniform reinjection probability.
\end{abstract}

\pacs{PACS numbers: 05.45.+b, 07.50.Ek}

\narrowtext

Intermittency characterized by the appearance of intermittent short chaotic
bursts between quite long quasiregular (laminar) periods is one of the critical
phenomena that can be readily observed in nonlinear dynamic systems. The
phenomenon was initially classified into three types according to the local
Poincar$\acute{e}$ map (type-I, II, and III) by Pomeau and Manneville \cite{1}.
The local Poincar$\acute{e}$ maps of type-I, II, and III
intermittencies are $y_{n+1} = y_n + ay_n^2 + \epsilon ~( a,~\epsilon > 0)$,
$y_{n+1} = ( 1 + \epsilon )y_n + ay_n^3 (a, ~\epsilon,~ y_n > 0)$, and
$y_{n+1} = -(1+\epsilon) y_n -ay_n^3 ( \epsilon, ~a > 0)$, respectively
\cite{2}.
The characteristic relation of type-I intermittency is $\langle l \rangle
\propto \epsilon^{-1/2}$, where $\langle l \rangle$ is the average laminar
length and $\epsilon$ is the channel width between the diagonal and the local
Poincar$\acute{e}$ map. Those of type-II and III intermittencies are
$ \langle l \rangle \propto \ln (1/\epsilon)$ where $1+\epsilon$ is the slope
of the local Poincar$\acute{e}$ map around the tangent point under the
assumption of uniform reinjection probability distribution (RPD). On the other
hand, some monographs \cite{6} suggested that the standard scaling should be
$\langle l \rangle \propto 1/\epsilon$.

Recently, however, it was found that the reinjection mechanism is another
important factor of the scaling property of the intermittency. In the
case of type-I intermittency, various characteristic relations appear dependent
on the RPD for the given local Poincar$\acute{e}$ map, such as $-\ln \epsilon$
and $\epsilon^{-\nu} ~(0\leq \nu \leq 1/2)$. When the lower bounds of the
reinjection (LBR) are below and above the tangent point the critical exponent
is always $-1/2$ and $0$, respectively, irrespective of the RPD. However when
the LBR
is at the tangent point the characteristic relations have various critical
exponents dependent on the RPD, such that when the RPDs are uniform, fixed and
of the form $x^{-1/2}$, the characteristic relations are $-\ln \epsilon,
~\epsilon^{-1/2}$, and $\epsilon^{-1/4}$ respectively \cite{3,4}.
In the case of type-II and III intermittencies, the characteristic relations
also have various critical exponents for a given local Poincar$\acute{e}$ map
such as $\epsilon^{-\nu} ~(1/2\leq \nu \leq 1)$ dependent on the RPD. When
RPDs are uniform, of the form $x^{-1/2}$ around the tangent point, and fixed
very close to the tangent point, the characteristic relations are
$\epsilon^{-1/2}$, $\epsilon^{-3/4}$, and $\epsilon^{-1}$, respectively
\cite{5}. In this report we discuss the characteristic relations of type-III
intermittency analytically, and obtain $\epsilon^{-1/2}$ characteristic
relation experimentally in an electronic circuit that consists of inductor,
resistor, and diode with uniform RPD.

 Since the local Poincar$\acute{e}$ map of type-III intermittency can be
described to be
$y_{n+2} = ( 1 + 2 \epsilon ){y_n} + b {y_n}^3$, which is the same as that of
type-II intermittency \cite{6}, it is enough to discuss the characteristic
relations
of type-II intermittency according to the RPD without loss of generality.
For the given local Poincar$\acute{e}$ map of type-II intermittency,
if we set a gate such that $| y_{in} | \leq c$ on deviations
in the laminar region, the laminar length $l\it (y_{in}, c)$ for the
reinjection at $y_{in}$ is obtained in the long laminar length approximation
\begin{equation}
l\it (y_{in},c) = {{2 \ln \left[{c \over y_{in}}\right] - \ln \left[{{a{c}^2 + \epsilon} \over {a{y_{in}}^2 + \epsilon}}\right]} \over {2 \epsilon}}.
\end{equation}
This is the result obtained in Ref. \cite{6} without considerartion of RPD.
Here if we consider a normalized RPD $P(y_{in})$ the average laminar length
$\langle l\it \rangle$ is given by
\begin{equation}
\langle l\it \rangle = \int^c_{\Delta} l\it(y_{in}, c)P(y_{in})d y_{in}
\end{equation}
where $\Delta$ is the value of $y_{in}$ representing the LBR.

We first consider the case of uniform RPD of the form ${1 / (c-\Delta)}$.
In this case the average laminar length is given by
\begin{eqnarray}
\langle l\it \rangle =
& &{} {\tan^{-1}\left(c \sqrt{a \over \epsilon}\right) - \tan^{-1}\left(\Delta \sqrt{a \over \epsilon}\right)} \over {\sqrt{\epsilon a} (c-\Delta)}\nonumber\\
& &{} - {{2 \Delta \ln \left(c \over \Delta\right) - \Delta\ln\left({a {c}^2 + \epsilon} \over {a \Delta^2 + \epsilon} \right)} \over {2 \epsilon (c - \Delta)}} .
\end{eqnarray}
In this equation if $\Delta$ is very close to the tangent point, that is
$\Delta^2 \ll \epsilon$, 
the second term is negligible in the limit $\epsilon \rightarrow 0$ because of
the factor $\Delta$ in the numerator and then
the characteristic relation is $\langle l\it \rangle \propto \epsilon^{-1/2}$.
However, when $\Delta$ is within the gate and not close to the tangent point,
the second term is still negligible in the limit $\epsilon \rightarrow 0$. Then
the characteristic relation is determined by the first term alone and is
a power type with critical exponent zero as
discussed in Ref. \cite{3}. What is interesting here is that these results are
different from those of Pomeau and Manneville's although the RPD is uniform.

We next consider the fixed RPD of which the form is $\delta(y_{in}-\Delta)$.
In this case the average laminar length is
\begin{equation}
\langle l\it \rangle = {{2 \ln \left(c \over \Delta\right) - \ln\left({a {c}^2 + \epsilon} \over {a \Delta^2 + \epsilon} \right)} \over {2 \epsilon}} .
\end{equation}
When $\Delta^2 \ll \epsilon$, the average laminar length can be
reduced to $\langle l \rangle \approx [\ln(\epsilon)-\ln(a\Delta^2)]/2\epsilon$
in the limit $\epsilon \rightarrow 0$. Then the characteristic relation is
$\langle l\it \rangle \propto \epsilon^{-1}$ since the constant
$\ln(a\Delta^2)$ is dominant over $\ln(\epsilon)$. Similarly when
$\Delta$ is far from the tangent point, the characteristic relation is
a power type whose critical exponent is zero.

In the above results since uniform and fixed RPDs are two extreme cases of
power type RPD, we
are able to give a general argument. Type-II
intermittency has power type characteristic relations such that
$\epsilon^{-\nu} (1/2 \leq \nu \leq 1)$ according to the RPD when
$\Delta^2 \ll \epsilon$ and has
a power form whose critical exponent is zero when $\Delta$ is not close to
the tangent point. 

To show this argument we further consider the case of nonuniform RPD of the
form
$1/(2 \sqrt{\Delta+c}\sqrt{y_{in}+\Delta})$, for which analytic calculation
of the characteristic relation is possible. The characteristic relation of
the average laminar length is
$\langle l\it \rangle \propto \epsilon ^{-3/4}$ in the limit
$\epsilon \rightarrow 0$ when $\Delta^2 \ll \epsilon$, and has
a power form whose critical exponent is zero when $\Delta$ is far from the
tangent point. These results show that type-II and III
intermittencies have various characteristic relations according to the RPD.

These characteristic relations of type-III intermittency were studied
experimentally in
the electronic circuit consisting of inductors, resistors,
and diodes shown in Fig. 1. A 1N4007 silicon junction diode and a 100 mH
inductor
(165 $\Omega$ dc resistance) connected in series were forced by a function
generator and a series connected 100 mH inductor and a 1N4007 diode are
connected in parallel with the first inductor. In this kind of electronic
circuit the characteristic relations of type-I intermittency were observed by
varying the amplitude or the offset of the external force \cite{8}.
In this circuit the amplitude of the external force can be varied by
multiplying
a sinusoidal signal from a function generator to a dc voltage from a
digital-analog converter using a multiplier (MPY100), and the dc voltage is
controlled by a personal computer. This apparatus can tune the amplitude
very precisely in the limit of noise from electronic elements. The frquency
and the bias voltage were fixed at 30 kHz and 0.4 volt, respectively. All the
external forces were added by using an operational amplifier,
and the noise from the power sources were reduced using by-pass capacitors.

The rectified voltages across the second diode were measured and each rectified
pulse was integrated using an integrating circuit, to obtain the experimental
data. Before integration 0.6 volt dc voltage was added to
the rectified pulses because the voltage drop across the
silicon diode was $-0.6$ volt. And after the adding, the
rectified voltages were reduced using a variable resistor, to prevent
distortion due to
the peak of the integrated voltage being higher than 15 volts.
Throughout the experiment we checked that the peaks of the rectified pulses
corresponded to those of the integrated pulses. The peaks of integrated
pulses were stored in the 40 M byte memory of the pentium computer by using
expanded memory manager via a 12-bit analog-digital converter.
All the systems were synchronized one another. The digitized time of the
analog-digital converter was 12 $\mu$sec. The digitized value of $\pm$2048
corresponds to $\pm$5 V, respectively. The chaotic outputs of the rectified
and integrated pulses were
also monitored, by using a digital storage oscilloscope (LeCroy 9310).

In the circuit various transitions from chaotic bands to stable fixed points
(or vice versa) were observed when the dc voltage from the computer was varied,
because of the nonlinear capacitance of the junction diode \cite{9}. To show
the temporal behavior of type-III intermittency in this system, we have
obtained them around the tangent bifurcation point near period-2 window as
shown in Fig. 2. Figure 2 (a) and (b) show long and short laminar phases
from when the amplitude of the external force was about $V=8.0V$ and
$8.01V$, respectively. The shapes are the typical temporal behaviors of
type-III intermittency. The continuously increasing and decreasing amplitudes
alternate, and are interrupted by the chaotic bursts.

To show more clearly that the temporal behaviors are type-III intermittency,
the $x_n$ vs $x_{n+2}$ return maps of the temporal behaviors are obtained as
given in Fig. 3 (a). In the figure, lines I and II are the return maps of
Fig. 2 (a) and (b), respectively.
Figure 3 (b) is $x_n$ vs $x_{n+4}$ return map that implies the local
Poincar$\acute{e}$ map of type-III intermittency can be expressed as that of
type-II. In the figures, the return maps of type-III are continuous near the
tangent point. This means that the LBR is very close to the tangent point.
Also the figures show the differences in slope around the tangent point,
which are related to the average laminar length. To confirm that Fig. 3 (b) is
the typical local Poincar$\acute{e}$ map of type-II intermittency, the return
maps near the tangent point are fitted with the cubic function,
$y_{n+1}=(1+\epsilon)y_n + ay_n^3 $. The parameters of lines I and II are
$a=1.2 \times 10^{-5}$
together, $\epsilon=0.050$ and $\epsilon=0.083$, respectively. This means that
the change in the external amplitude causes the changes in the slope of the
local Poincar$\acute{e}$ map.

To obtain the RPD from the return map, the reinjecting region, between
$V_{t} = 1.31V$ and $1.67V$, is divided into 150 sections and the number of
reinjections at each section is counted. Figure 4 is a log-log plot of the
total number of the reinjection at each section, where $V_r$ is the voltage of
reinjections and $V_{t}$ is the voltage of the tangent point.
As given in the figure, the slope of the RPD turns out close to zero
if the gate size is small, which means that the RPD is uniform.

In the intermittency region, the average laminar
lengths were obtained by varying the amplitude of the external force.
In this measurement, we reduced the voltage from digital-analog converter
by a factor of $50$ using resistors connected in series. This was done to
enable a fine tuning of $\epsilon$ after the addition of a further dc voltage
to bring the amplitude of the
forcing signal near to the bifurcation point. The step size of the voltage
from the digital-analog converter was about 0.05 mV. As the voltage
reduced, the length of the laminar was counted and when the longest laminar was
less than $10^4$, the computer began storing the data. We assume that
the last point at which the length of the longest laminar is larger than $10^4$
is the bifurcation point. In the experiment, if the length of the longest
laminar was larger than $10^4$, we could not observe chaotic bursts between
regular periods. Also, at each voltage, $10^4$ laminar phases were obtained and
300 steps of total step size were varied, which corresponds to about 15 mV of
total variation. Figure 5 is a log-log plot of the average laminar lengths vs
$\epsilon$ where $\epsilon = |V - V_t|$. In the figure the dots are
experimental results and the solid lines are the fitting of the experimental
data. The slope of the solid lines are the critical exponents of the
characteristic relations. Line A clearly shows that the characteristic
relation is $ \langle l \rangle \propto {|V - V_t |}^{-\nu}$ with approximately
$\nu \approx 0.5$. The critical exponent is remarkably similar to the result
obtained theoretically with the LBR very close to the tangent point.

We also obtained the critical exponent of zero with LBR above the tangent
point, labelled by line B in Fig. 5. The line was obtained when the external
force was around $12.4V$ and bias voltage is $0.8V$. In this region hysterisis
crisis appears \cite{10} and the stable fixed point is divided into two
chaotic bands (one is above and the other below the tangent point) after
the tangent bifurcation. This means the LBR is far from the tangent point.
To recapitulate, if LBR is far from the tangent point, the critical exponent
is zero irrespective of RPD. Line B clearly shows constant laminar lengths
when $\epsilon$ is small. The result again agrees well with the theoretical
predictions for the LBR far from the tangent point. The two lines clearly show
the characteristic relations due to the RPD in experiment.

In summary, we have obtained various characteristic relations of type-II and
III intermittencies such as $\langle l\it \rangle \propto \epsilon^{-\nu}$
$(1/2 \leq \nu \leq 1) $ depending on the RPDs when the LBR is very close to
the tangent point. Also zero critical exponent is obtained irrespective of the
RPD when the LBR is far from the tangent point. In an inductor-resistor-diode
circuit, these characteristic relations were studied in experiment. The RPD of
type-III intermittency appearing in this circuit is uniform and the local
Poincar$\acute{e}$ map can be replaced with type-II intermittency whose local
Poincar$\acute{e}$ map is of the form
$y_{n+1} = (1+ \epsilon) y_n + a y_{n}^3 $. Thus, $-1/2$ and $0$ critical
exponents are obtained when the LBR is very close to and far from the
tangent point, respectively. These experimental results well agree
with those of the theoretical analysis.

This research was supported in part by the Ministry of Science and Technology
of Korea
under the Project ``High-Performance Computing-Computational Science and
Technology (HPC-COSE)'' and by the Ministry of Education of Korea, Project
No. BSRI-97-2414. We thank Dr. J. M. Smith for useful discussions and
suggestions.

\begin{figure}

\epsfig{file=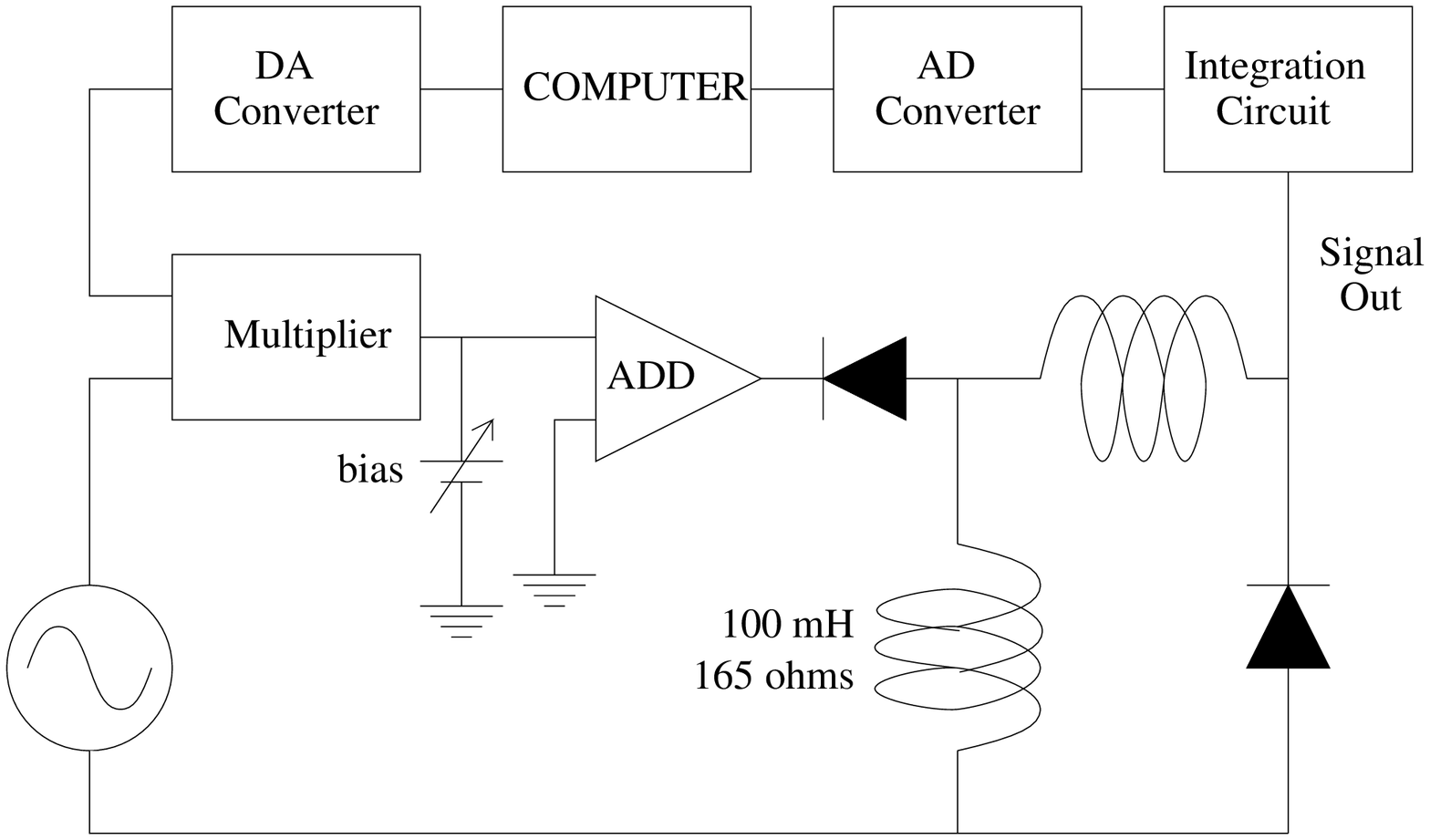,width=.4\textwidth}
\caption{Schematic diagram of experimental setup.}

\epsfig{file=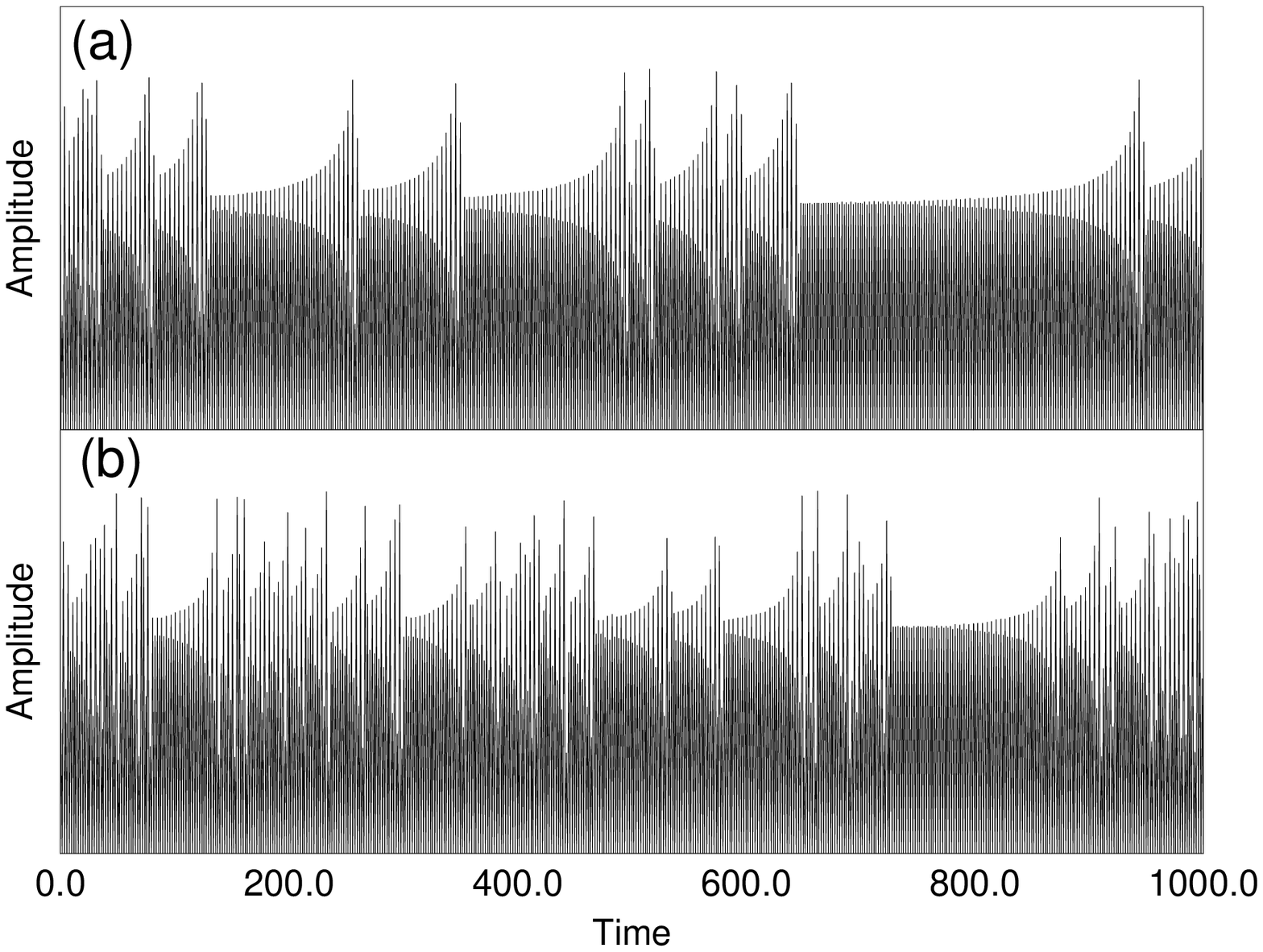,width=.4\textwidth}
\caption{Temporal behaviors of intermittency near period-2 window for
(a) long and (b) short laminar phases.}

\epsfig{file=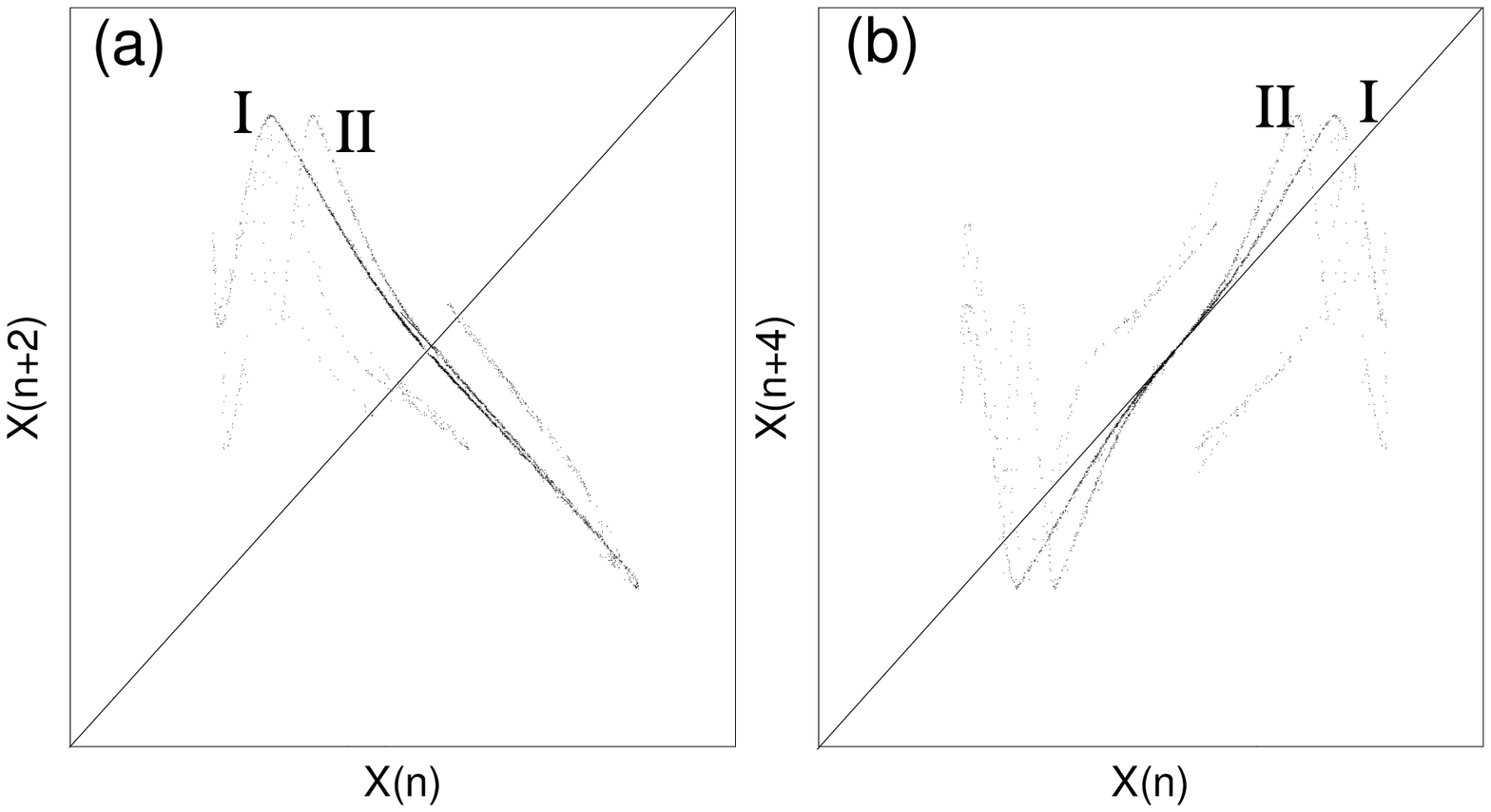,width=.4\textwidth}
\caption{(a) $x_n$ vs $x_{n+2}$ and (b) $x_n$ vs $x_{n+4}$ return maps near
period-2 window. Line I and II are the maps of Fig. 2 (a) and (b).}

\epsfig{file=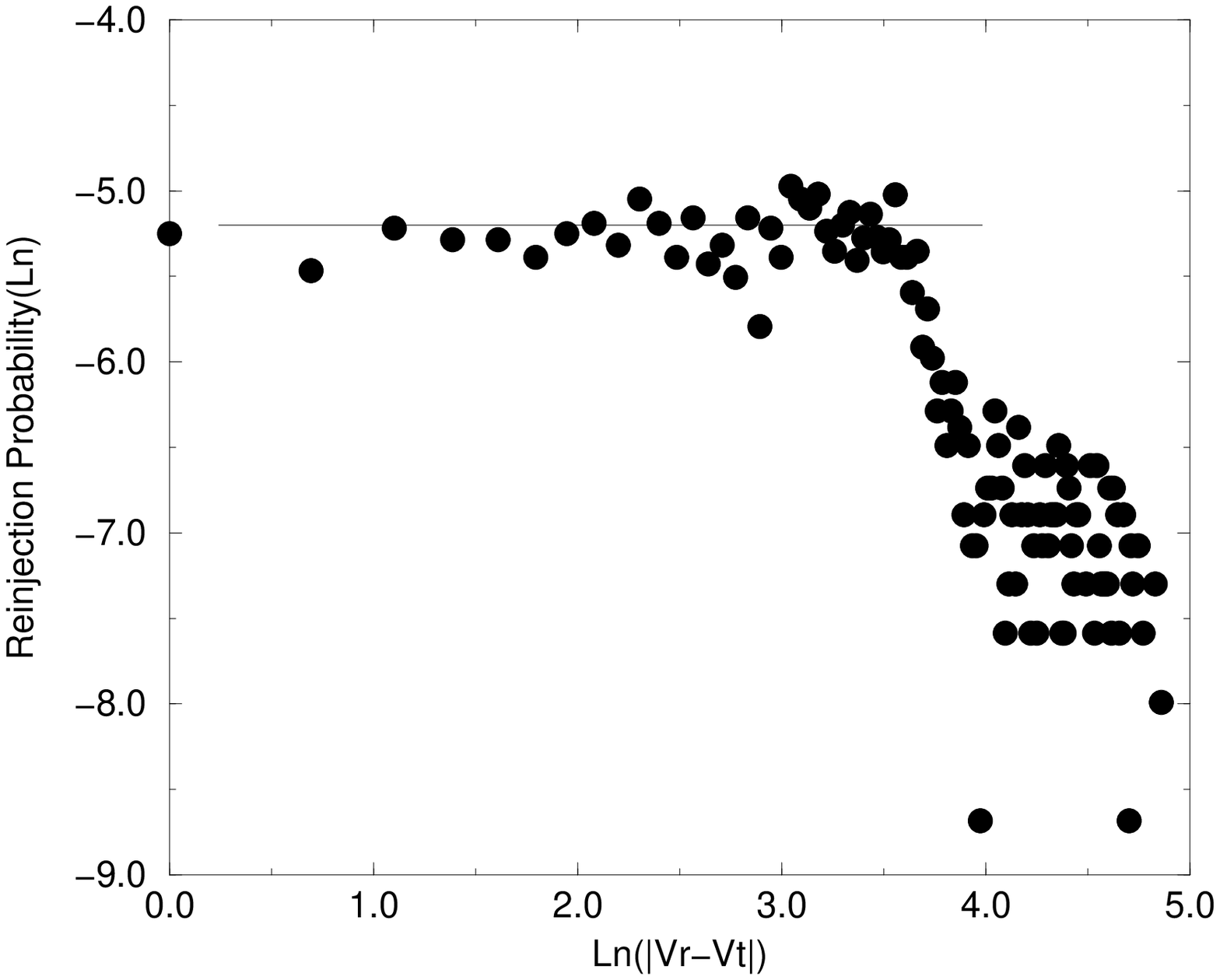,width=.4\textwidth}
\caption{RPD near period 2-window. The dots are experimental data and the
solid line is the fitting of the data when the gate size is small. The slope
of the solid line is 0, so the RPD is approximately uniform.}

\epsfig{file=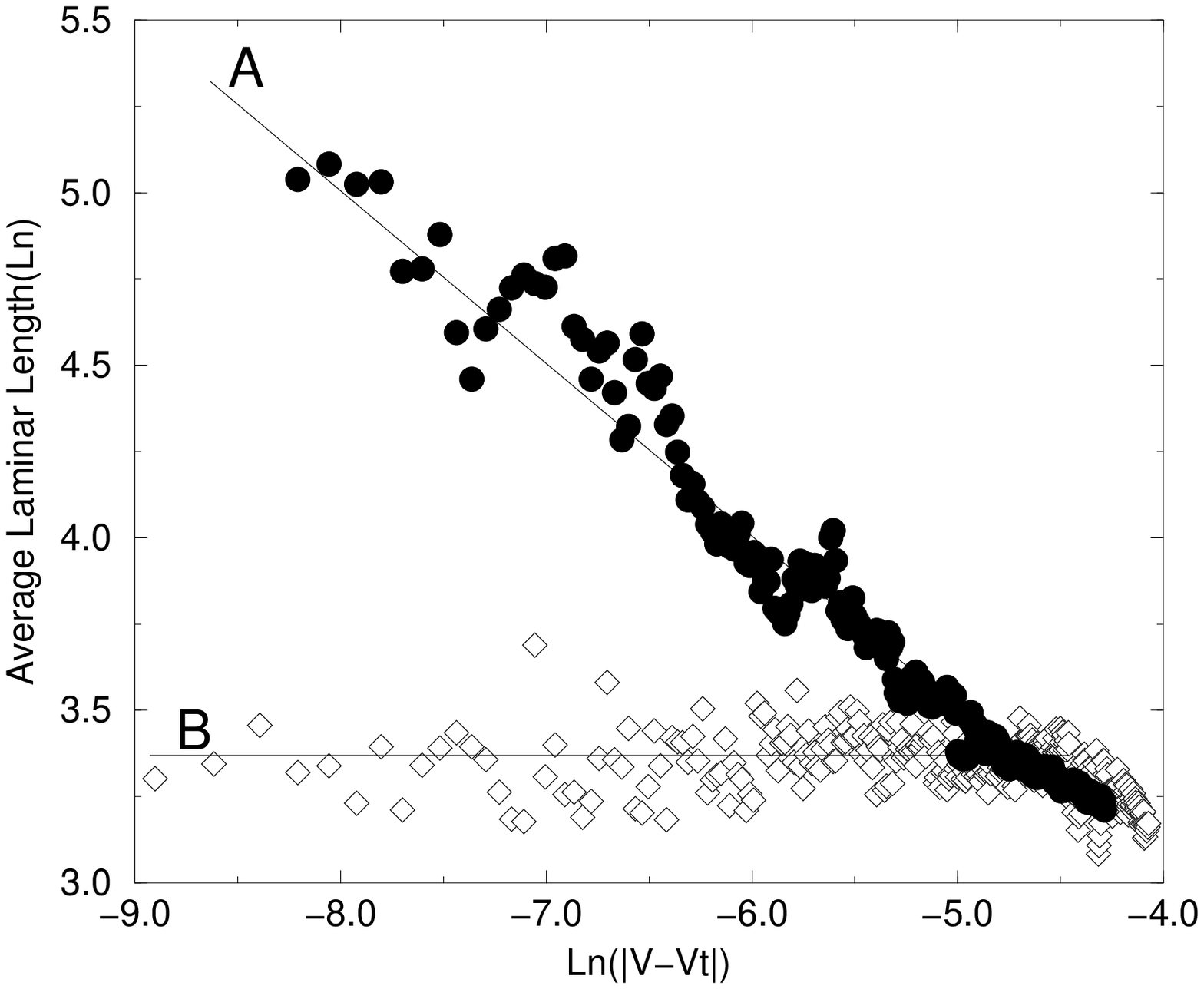,width=.4\textwidth}
\caption{The average laminar lengths vs $|V-V_t |$. The figure shows well that
the critical exponents are $-$1/2, and 0, when the LBRs are very close to
(line A) and above (line B) the tangent point, respectively. The dots are
experimental data and the solid lines are the fitting of the data with the
slopes of $-$1/2 and 0, respectively.}

\end{figure}


\begin{references}
\bibitem{2} The typical Poincar$\acute{e}$ map of type-II intermittency
composes of $\theta_{n+1} = \theta_n + \Omega$ as well as
$y_{n+1} = (1+\epsilon)y_n +a y_n^3$. However the radius variation alone can
determine the average laminar length and consequently the characteristic
relation.
\end{references}
\end{document}